\documentclass[conference]{IEEEtran}
\IEEEoverridecommandlockouts

\usepackage{cite}
\usepackage{amsmath,amssymb,amsfonts}
\usepackage{algorithmic}
\usepackage{algorithm}
\usepackage{graphicx}
\usepackage{textcomp}
\usepackage{xcolor}
\usepackage{comment}
\def\BibTeX{{\rm B\kern-.05em{\sc i\kern-.025em b}\kern-.08em
    T\kern-.1667em\lower.7ex\hbox{E}\kern-.125emX}}
\begin{document}

\title{Efficient Transmission of Radiomaps via Physics-Enhanced Semantic Communications\\
\thanks{This work was supported in part by the National Science Foundation under Grants No. 2413622, No. 2231209, No. 2315596, No. 2244219, No. 2146497, No. 2431452; and in part by a grant from BoRSF under contract LEQSF(2024-27)-RD-B-03. (Corresponding author: Songyang Zhang.)}
}

\author{\IEEEauthorblockN{Yueling Zhou$^*$, Achintha Wijesinghe$^\star$, Yue Wang$^\dagger$, Songyang Zhang$^*$, Zhipeng Cai$^\dagger$}
\IEEEauthorblockA{$^*$University of Louisiana at Lafayette, Lafayette, LA, USA, 70504\\$^\star$University of California at Davis,  Davis, CA, USA, 95616 \\$^\dagger$Georgia State University, Atlanta, GA, USA, 30302}}

\maketitle

\begin{abstract}
Enriching information of spectrum coverage, radiomap plays an important role in many wireless communication applications, 
such as resource allocation and network optimization. To enable real-time, distributed  spectrum management, particularly in the scenarios with unstable and dynamic environments, the efficient transmission of spectrum coverage information for radiomaps from edge devices to the central server emerges as a critical problem. In this work, we propose an innovative physics-enhanced semantic communication framework tailored for efficient radiomap transmission based on generative learning models. Specifically, instead of bit-wise message passing, we only transmit the key ``semantics" in radiomaps characterized by the radio propagation behavior and surrounding environments, where 
semantic compression schemes are utilized to reduce the communication overhead. Incorporating the novel concepts of Radio Depth Maps, the radiomaps are reconstructed from the delivered semantic information backboned on the conditional generative adversarial networks. 
Our framework is further extended to facilitate its implementation in the scenarios of multi-user edge computing, by integrating with federated learning for collaborative model training while preserving the data privacy. 
Experimental results show 
that 
our approach achieves high accuracy in radio coverage information recovery at ultra-high bandwidth efficiency, 
which has great potentials in many wireless-generated data transmission applications.
\end{abstract}


\section{Introduction}
Next-generation wireless 
technologies, such as vehicle-to-everything (V2X) and integrated sensing and communications (ISAC), have experienced substantial advancements, leading to increasing demand of high-precision radiomaps
for panoramic awareness  
of the radio frequency (RF) spectrum coverage and radio environments \cite{Bi2019}.
Characterizing the multi-dimensional distribution of RF power spectral density (PSD) across various locations, frequencies, and time, radiomap plays a vital role in spectrum management tasks, such as unmanned aerial vehicle (UAV) path planning \cite{ZhangUAV2021} and three-dimensional (3D) environment reconstruction \cite{Zeng2022}. Usually,
dense radiomaps are reconstructed from sparse observations collected by mobile sensors and user devices \cite{Romero2022}, making radiomap estimation (RME) a critical challenge in modern wireless communications. However, most existing radiomap-assisted applications focus on RME with a global view at single snapshot, 
while ignoring the fact that sparse observations are collected by edge devices or driving test on-the-fly, and are then transmitted to the central server for data processing \cite{RMEGAN2023}. 
For example, in the drone-assisted emergency communications, UAVs 
collect PSD observations for 3D environment reconstruction and temporary base station deployment, where timely radiomap transmission 
boosts the efficiency of spectrum management~\cite{Wu2020}.
In this sense, efficient transmissions of radiomaps from edge devices allow RME and radiomap-assisted network optimization in dynamic and complex spectrum environments.

\begin{figure}[t]
    \centering
    \includegraphics[width=0.9\linewidth]{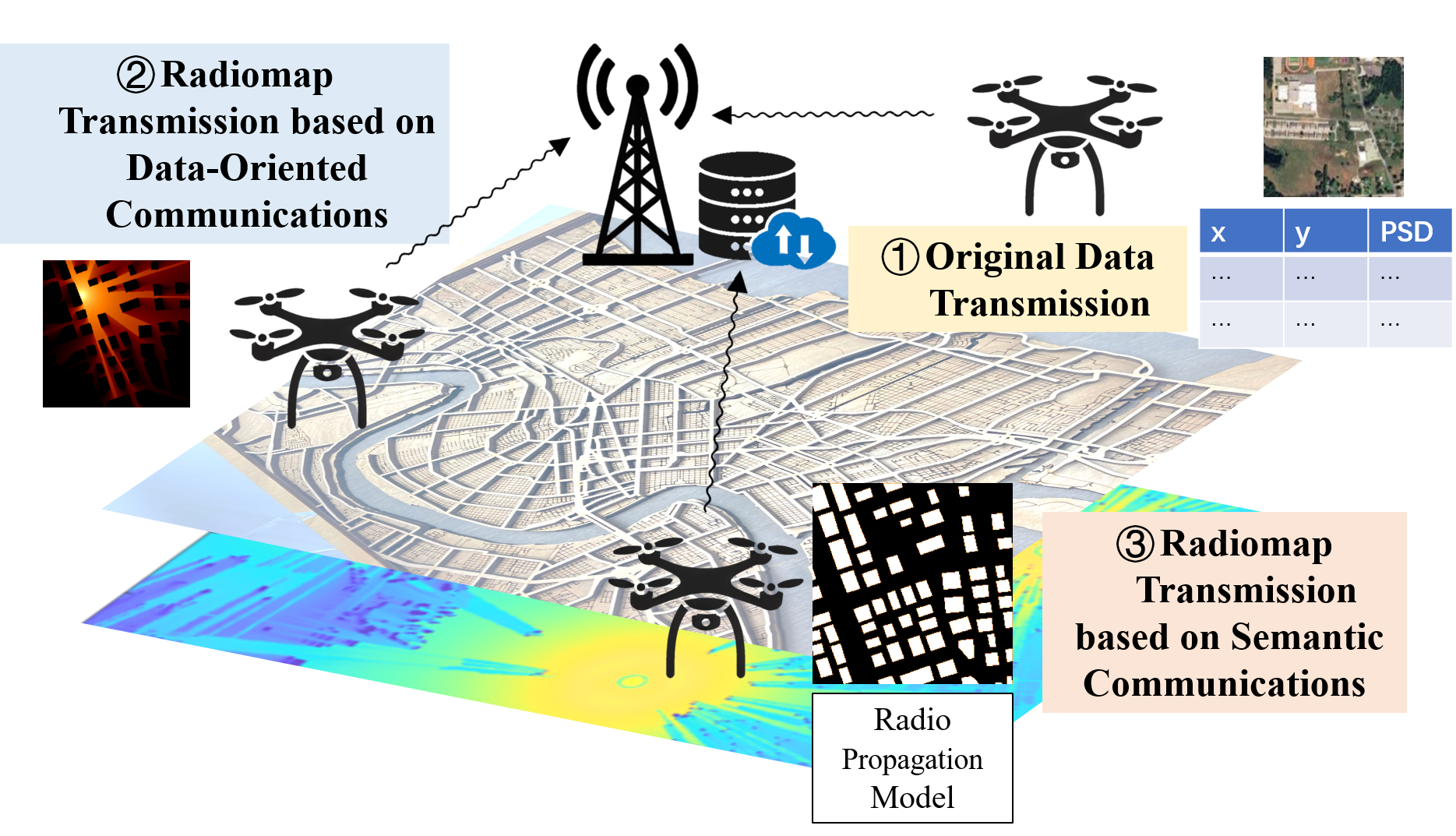}
    \caption{Three types of radiomap transmission methods: 1) Type 1 is the original data transmission, which transmit the landscape image together with the PSD value and coordinates of each sensing point. 2) Type 2 is data-oriented communications, which first reconstruct the radiomaps at the edge devices and utilize deep learning models for JSCC. 3) Type 3 is our proposed method via semantic communications,  which only transmits the most critical semantic information for radiomap reconstruction at the receiver end.}
    \label{fig:1}
    \vspace{-5mm}
\end{figure}

 \begin{figure*}[t]
    \centering
    \includegraphics[width=0.8\linewidth]{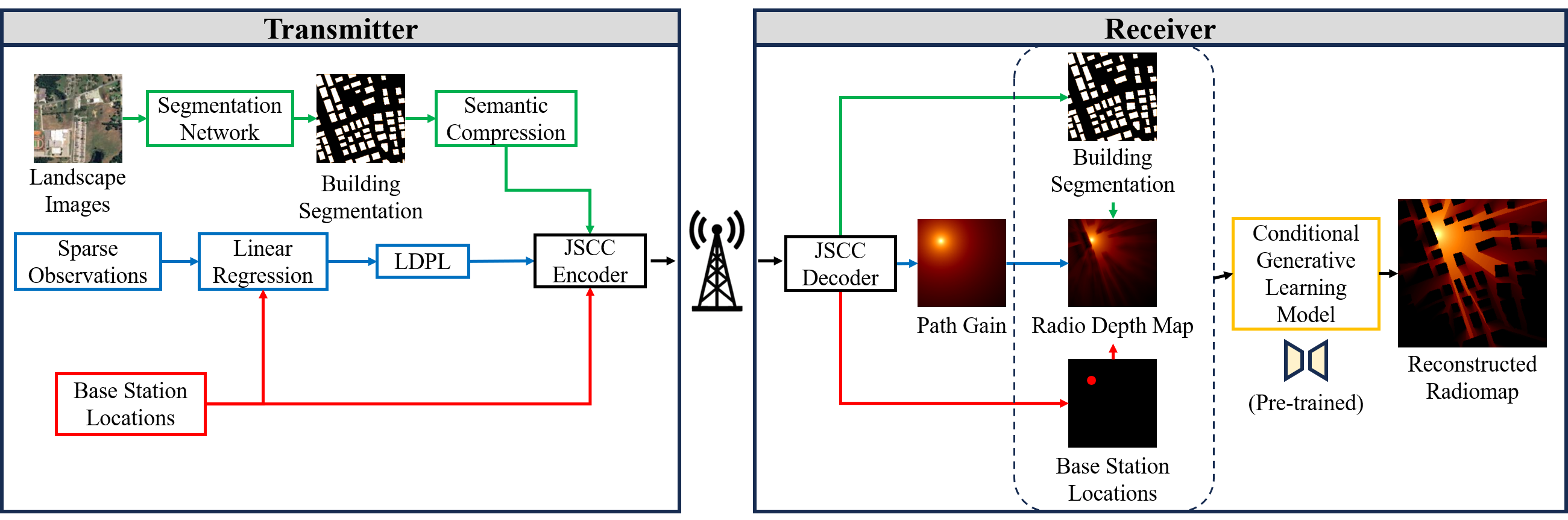}\vspace{-2mm}
    \caption{Proposed semantic communications for radiomap transmission (inference phase): 1) At the transmitter side, raw data including the landscape images and sparse PSD samples are collected by edge devices, which are then encoded into key ``semantics" characterized by building segmentation, radio propagation models and base station locations. With semantic compression and additional JSCC, these features are transmitted to the receiver side for radiomap reconstruction; 2) At the receiver side, the decoded semantic features, together with the physics-processed radio depth map are inputted into the pre-trained cGAN for radiomap reconstruction, which can be further utilized for downstream radiomap-assisted spectrum management, such as outage map prediction.}
    \label{fig:2}
    \vspace{-3mm}
\end{figure*}

Traditional spectrum information transmission relies on the data-oriented communications \cite{YangDO2019}, which focus on the bit-wise message delivery. As shown in Fig.~\ref{fig:1}, one type of transmission methods is the direct transmission of original sparse observations collected by mobile devices, together with the landscape information such as satellite images. 
Then, the RME is conducted at the receiver end (cloud server) for downstream spectrum management. However, due to the ignorance of correlations among different sparse samples and the high resolutions of auxiliary landscape data, the raw data sharing might result in heavy communication overheads, 
particularly in the resource-limited sensor networks. On the other hand, recent development of learning techniques 
enables RME 
in the edge device, after which 
the learning-based joint source and channel coding (JSCC) is utilized for radiomap transmission~\cite{Bourtsoulatze2019,Saidutta2021}, e.g. type-2 shown in Fig.~\ref{fig:1}. Despite some successes in improving communication efficiency, there are still 
redundant information in radiomap transmission. For example, the PSD observations in different locations 
usually share 
similar radio propagation behavior within a given sensing region of the edge devices. How to efficiently transmit  radiomaps in real-time, distributed spectrum management remain an open question.

Recent advancements in semantic communications, as a new era to transmit information, motivate us to seek 
for 
efficient radiomap transmission in the dynamic spectrum environments. 
Instead of focusing on the bit-wise data transmission and recovery as in conventional data-oriented communications, semantic communications aim to convey the most critical ``semantic" information for the accomplishment of downstream tasks \cite{Luo2022}. Existing semantic communication frameworks usually leverage 
a straightforward use of deep learning techniques, where the semantic features are extracted in the transmitter side and 
then delivered to the receiver end for data reconstruction. Typical examples include those based on auto-encoders \cite{Weng2023}, transformers \cite{Zhou2022} and diffusion models \cite{Wijesinghe2024}. However, most existing semantic communication frameworks are designed for handling the human-generated contents, while few of them deal with the wireless-generated data. Different from the human-generated contents, the semantics of wireless-generated data shall be related to both physics models (e.g., radio propagation in RME) and auxiliary data (e.g., images). To the best of our knowledge, it is still a lack of research results to efficiently embed the physical model information with semantics communications in the literature.

To fill the gap, 
we propose a novel physics-enhanced semantic communications for efficient transmission of radiomaps, illustrated as type-3 in Fig.~\ref{fig:1}. 
Specifically, we first extract the key semantic information and predict the radio propagation models from sparse observation samples to characterize the radio propagation behavior. Then, 
different semantic compression schemes are explored for bandwidth compression. With a pre-trained conditional generative adversarial nets (cGAN)~\cite{Mirza2014} shared at both transmitter and receiver, the dense radiomap can be reconstructed at the receiver end. Moreover, 
for privacy preserving in multi-user semantic communications, federated learning is utilized 
for collaborative model training. Our contributions are summarized as follows:

\begin{enumerate}
    \item We are the first to introduce the 
    physics-enhanced semantic communications for radiomap transmission, which significantly reduce the communication overhead and increase the accuracy of RME. 
    \item Thanks to the effective feature exploration in learning-based embedding, we provide comprehensive designs for semantic compression to dramatically boost 
    communication efficiency.
    \item To enable multi-user semantic communications based on generative learning models, federated cGAN (FedCGAN) is leveraged 
    for collaborative model training 
    while keeping local data private.
    \item Our approach is verified through extensive experiments for both RME and downstream spectrum management tasks, which achieves efficient and accurate radiomap transmission across diverse communication scenarios.
\end{enumerate}
Note that, to avoid confusion, we shall use the terminology of ``transmitter" and ``receiver" for semantic communications, and use ``base stations" for RME unless stated otherwise.

\section{Single-User Radiomap Transmission}
We start to introduce our physics-enhanced semantic communications for radiomap transmission in a single-user case. 

\subsection{Overall Architecture} \label{backbone}
\begin{figure}
    \centering
    \includegraphics[width=0.8\linewidth]{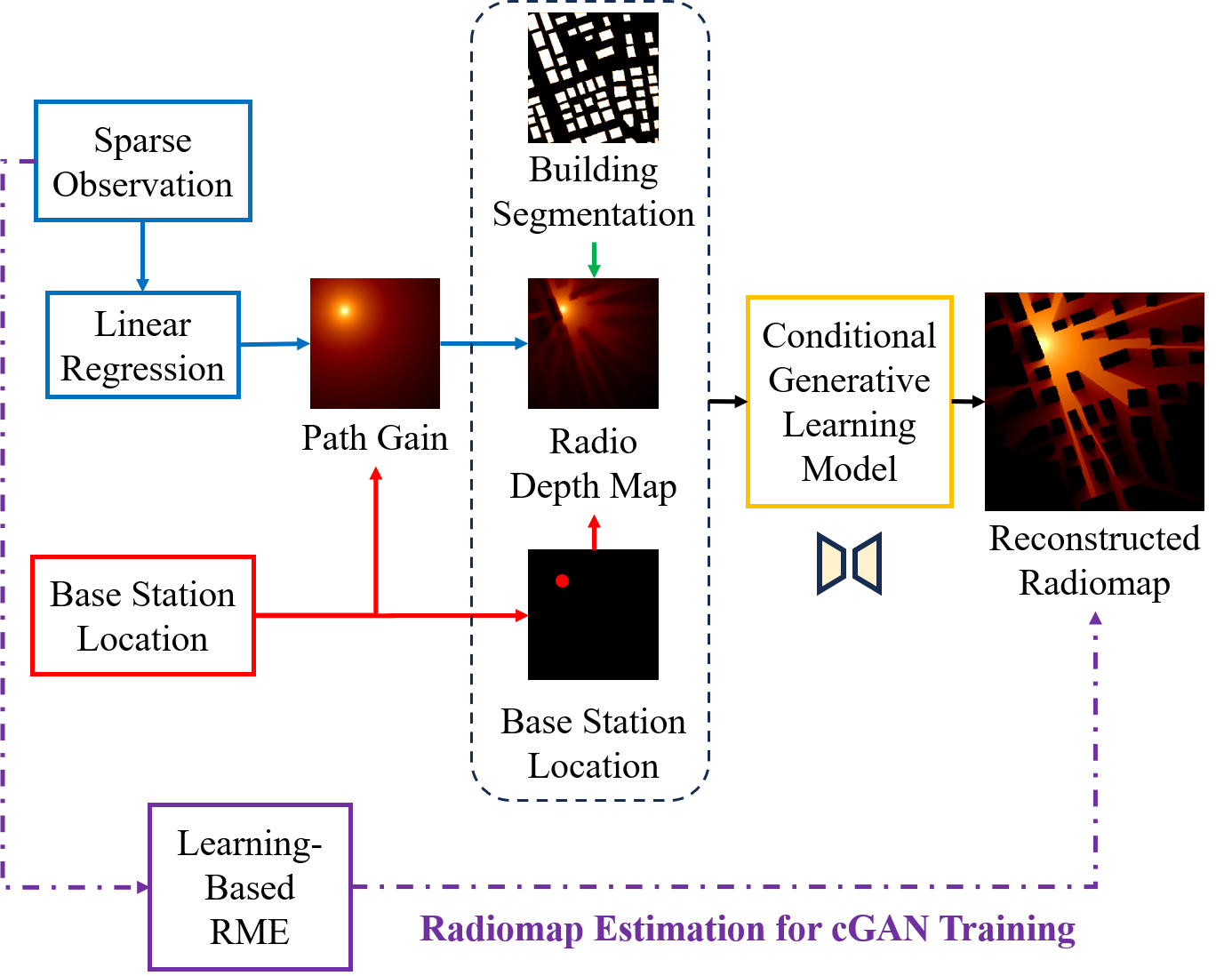}
    \caption{Training phase of cGAN backbone: dense radiomaps for training can be generated by existing learning-based RME from the sparse observations.}
    \label{fig:train}
    \vspace{-4mm}
\end{figure}

In the semantic communications of radiomap transmission, the key semantic information shall be able to characterize the radio propagation behavior. 
In our proposed framework as in Fig.~\ref{fig:2}, the mobile devices (e.g., UAVs and smart phones) 
collect the sparse PSD samples in a given region, together with the landscape images via cameras. 
They are further embedded by building segmentation, radio propagation model parameters and base station locations. 
Given these semantics, a cGAN can be pre-trained for radiomap reconstruction at the receiver side.
The training and inference phases of our physics-enhanced semantic communication framework are 
illustrated as in Fig.~\ref{fig:train} and Fig.~\ref{fig:2}, respectively. 
\subsubsection{Training Phase}
In our framework, the reconstruction of radio coverage information is based upon a pre-trained cGAN as shown in Fig. \ref{fig:train}, which takes the building segmentation, radio depth map, and base station locations as input. The building segmentation can be estimated from the collected landscape images through existing learning-based segmentation neural networks, such as Unet and ResNet \cite{Minaee2022}. The input radio depth map and output radiomap of cGAN in the training sets can be generated from the sparse PSD observations~\cite{RMEGAN2023}. Specifically, the radiomap estimation for cGAN training can be implemented via the state-of-the-art (SOTA) RME methods, such as RME-GAN \cite{RMEGAN2023} or RadioUnet \cite{Levie2021}. The generation of radio depth map will be illustrated in detail later in Section \ref{RDM}, with the parameters of log-distance path loss (LDPL) model estimated by multi-variable linear regression. After training, the cGAN is deployed at the receiver end for the radiomap reconstruction in the inference phase.

\subsubsection{Inference Phase} 
The real-time collected sparse observations and landscape images are processed into LDPL model parameters and building segmentation images 
in the inference phase. Together with the base station locations, the building segmentation and LDPL model parameters are encoded by 
available JSCC blocks, which are then transmitted to the receiver side. Upon receiving and decoding the delivered semantics, the radio depth map is generated from LDPL models and building information. The base station location is encoded in a binary images, with $1$ at base station location; otherwise, $0$ for the pixel value. Finally, 
inputting the radio depth map, building segmentation, and base station location into the pre-trained cGAN, radiomaps can be reconstructed for downstream tasks.

The details of each block in our physics-enhanced semantic communications will be illustrated in the following sections.

\subsection{Prediction of LDPL Model Parameters}
We describe the radio propagation model by the LDPL model, which serves as one of the key component to generate the physics-enhanced radio depth map. Particularly, 
the LDPL model is denoted as 
\begin{equation}
\text{PL}(d) = \text{PL}_0 + 10 \cdot \theta \cdot \log_{10}\left(\frac{d}{d_0}\right) + X_\sigma,
\label{eq:LDPL}
\end{equation}
where $\text{PL}(d)$ is the path loss at distance $d$, $\text{PL}_0$ is the path loss at a reference distance $d_0$, $d$ represents the distance between a certain location to the base station (BS), $d_0$ is the reference distance (usually 1 meter), $\theta$ is the path loss exponent, and $X_\sigma$ is a Gaussian random variable with zero mean and standard deviation $\sigma$. Combining $10\cdot \theta$ as $\tilde{\theta}$ and ignoring $X_\sigma$, the LDPL model in Eq.~\eqref{eq:LDPL} can be simplified as 
\begin{equation}\label{PL}
\text{PL}(d) = \text{PL}_0 + \tilde{\theta} \cdot \log_{10}\left(d\right)
,\quad \mbox{with }
d = ||p-p_0||^2
\end{equation}
where $p_0$ is the BS location and $p$ is the point of interest. Then, the parameters of LDPL models can be predicted via linear regression by minimizing the error between Eq. \eqref{PL} and sparse observations. To eliminate the impact from buildings, we restrict LDPL parameter estimation by samples within a specific radius around the transmitter, where signal decay is more representative of the natural fading pattern.

\vspace{-2mm}
\subsection{Radio Depth Map}\label{RDM}
To enhance the physics features with surrounding environmental information, we introduce the radio depth map (RDM) $\mathcal{X} = M_D \in \mathbb{R}^{N \times N}$ for feature extraction \cite{Zhang2024Mag}, calculated by
\begin{equation}\label{depthmap} M_D(x,y) = n\left(\sum_{t=1}^{N_T} \text{PL}_t(x,y) \cdot B_t(x,y)\right), \end{equation}
where $(x, y)$ represents the two-dimensional coordinates of the target location, $N_T$ is the number of BSs, $\text{PL}_t(x, y)$ (same as $\text{PL}(d)$ in Eq. \eqref{PL}) refers to the LDPL from the $t$-th transmitter, and $n(\cdot)$ is a max normalization function that scales the maximum value in the RDM to 1. The term $B_t(x, y)$ captures the influence of buildings from the viewpoint of the $t$-th transmitter.
Let $C(\ell, \ell_t)$ be the set of pixels in the direct line-of-sight path between the target location $\ell = (x, y)$ and the $t$-th transmitter's location $\ell_t = (x_t, y_t)$. Then, the building effect is measured by 
\begin{equation} B_t(x, y) = \frac{\sum_{i \in C(\ell, \ell_t)} (1 - M_U(i))}{\sum_{i \in C(\ell, \ell_t)} 1}, \end{equation}
where $B_t(x, y)$ represents the ratio of non-building pixels between $\ell$ and $\ell_t$, and $M_U$ denotes 
the building segmentation map, i.e., the building area is 1; otherwise are 0's. A higher value of $B_t(x, y)$ implies fewer obstacles along the propagation path, indicating a weaker 
shadowing effect.

\vspace{-2mm}
\subsection{Semantic Compression}
Next, we design the semantic compressions of building segmentation for relaxing the bandwidth requirement. 
\subsubsection{Semantic Compression with VQVAE}
We first leverage Vector Quantized Variational Autoencoder (VQVAE) \cite{VQVAE} to achieve semantic compression of building segmentation images. Specifically, we 
embed the building segmentation map \( M_U \) into a low-dimensional latent space through an encoder \( E(\cdot) \), where the latent representation is \( \mathcal{Z} {=} E(M_U) {\in} \mathbb{R}^{A \times B \times L} \), consisting of $AB$ latent components $\mathbf{z}=\{\mathbf{z}_1,\cdots,\mathbf{z}_{AB}\}$. Within this latent space, we construct a codebook \( \mathbf{W} = [\mathbf{w}_1, \mathbf{w}_2, \dots, \mathbf{w}_n] \in \mathbb{R}^{L \times n} \), where each codeword \( \mathbf{w}_i \in \mathbb{R}^L \) serves as a representative pattern of building structures.
For each component \( \mathbf{z_i} \) in the latent representation, we perform quantization by selecting the nearest codeword \( \mathbf{w}_{k_i} \) from \( \mathbf{W} \), via minimizing the Euclidean distance as
\[
k_i = \arg \min_j \| \mathbf{z_i} - \mathbf{w}_j \|_2^2.
\]
The index vector \( \mathbf{v} = [v_1, v_2, \dots, v_{AB}] \), where \( v_i = k_i \), records the indices of these selected codewords, allowing us to approximate \( \mathbf{z} \) with \( \mathbf{\hat{z}} = \{ \mathbf{w}_{k_1}, \mathbf{w}_{k_2}, \dots, \mathbf{w}_{k_{AB}} \} \). With the codebook pre-trained and shared in advance, we only need to transmit the index vector $\mathbf{v}$ for building segmentation reconstruction at low transmission cost.
Through the VQVAE decoder \( D(\cdot) \),  a quantized approximation of the original building segmentation map can be obtained via \( \hat{M}_U = D(\mathbf{\hat{z}}) \).
\subsubsection{Semantic Compression with JPEG}
In addition to VQVAE, we can also treat the binary segmentation map as images and apply JPEG compression to encode the building information. As a widely used lossy compression technique, JPEG can efficiently reduce the data size while preserving the essential structural features of the building segmentation. At the receiver side, a threshold filter is applied to ensure that the decoded building segmentation map remains in a binary format, where each pixel is either classified as building or non-building for radiomap reconstruction. 

\subsection{Conditional Generative Learning}
Finally, we establish our cGAN backbone, including 
two 
components: a generator $G$ and a discriminator $D$. In contrast to traditional GANs which generate data from random noise, the generator in a cGAN 
produces data based on specified features $\mathcal{F}$ and incorporates additional information $\mathbf{I}$ into the discriminator. In our framework, the features are formed by $\mathcal{F}=\{M_U, M_T, M_D, \mathbf{I}\}$, representing the building segmentation, BS location map and the radio depth map, 
as shown in the receiver side of Fig. \ref{fig:2}. The additional information $\mathbf{I}$ can be designed as a function of the region features or other task-oriented labels if any. We have the 
cGAN objective function: 
\begin{align}
    \min_G\max_D \, V(D,G) = \, & \, \mathbb{E}_{y,I\sim p_{\rm data}(\bf{y},\bf{I})}[\log D(\bf{y,I})] \, +\nonumber\\
    &\, \mathbb{E}_{f\sim p_{\rm data}(\bf{\mathcal{F}})}[\log (1-D(G(\bf{\mathcal{F}}),\mathbf{I}))],
\end{align}
where $p_{\rm data}(\mathbf{y},\mathbf{I})$ is the joint distribution with prior knowledge $\mathbf{I}$, and $\mathbf{y}$ is the generated data. The optimization of this min-max problem in cGAN is typically split into two sub-problems, with $D$ and $G$ trained iteratively. In this work, we optimize $D$ via $L_D=-V(D,G)$, and $G$ by 
\begin{equation}\label{CGAN}
L= \alpha L_G+ L_{MSE},
\end{equation}
where
   $ L_G=\mathbb{E}_{x\sim p_{\rm data}(\bf{\mathcal{F}})}[\log (1-D(G(\bf{\mathcal{F}}),\mathbf{I}))]$
and $L_{MSE}$ refers to the Mean Squared Error (MSE) of RME.

\section{Multi-User Semantic Communications}

\begin{figure}[t]
    \centering
    \includegraphics[width=0.85\linewidth]{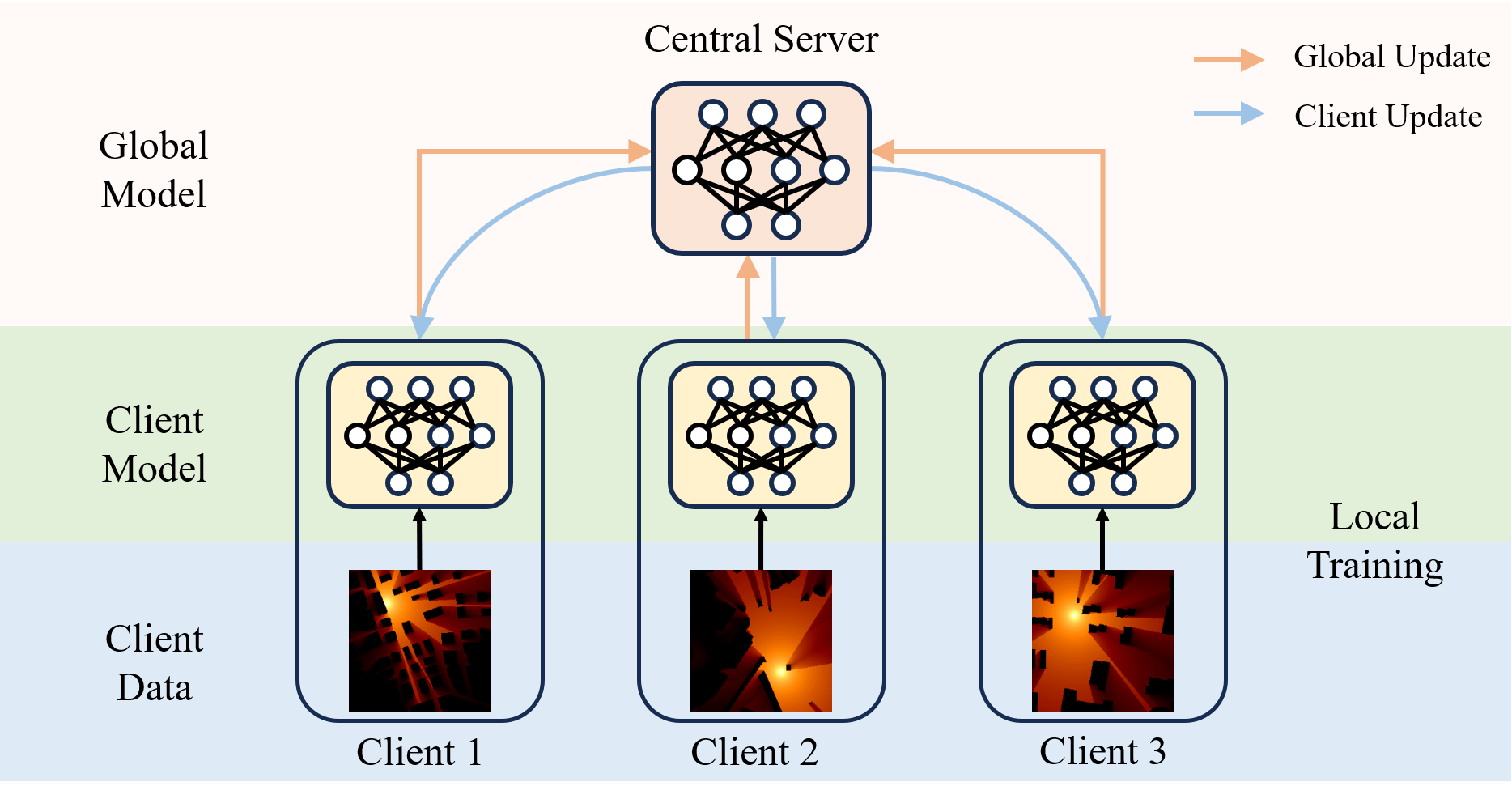}
    \caption{Federated learning scheme for multi-user semantic communications. Each client trains its own model to reconstruct the local radiomap. At each synchronization cycle, each client shares their local model parameters with the global server, which aggregates the updates and returns the improved global model parameters to the clients.}
    \label{fig:3}
\end{figure}
\begin{algorithm}[t]\label{fed_alg}
\caption{FedCGAN for Multi-user Scenarios}
\begin{algorithmic}[1]
\STATE Initialize global model parameters $W_{\text{global}}$
\FOR{each communication round $t = 1, \dots, T$}
    \STATE Sample a subset of clients $S_t$ from the total client pool
    \FOR{each client $i \in S_t$ \textbf{in parallel}}
        \STATE Send the global model $W_{\text{global}}$ to client $i$
        \STATE Client $i$ initializes local model parameters $W_i = W_{\text{global}}$
        \FOR{each local epoch $e = 1, \dots, E$}
            \FOR{each batch $b$ in client $i$'s local dataset}
                \STATE Compute gradients $\nabla W_i$ using local data batch $b$
                \STATE Update $W_i$ based on gradients $\nabla W_i$ (e.g., using SGD or Adam)
            \ENDFOR
        \ENDFOR
        \STATE Send updated local model parameters $W_i$ back to the server
    \ENDFOR
    \STATE Aggregate $W_{\text{global}} = \text{Aggregate}(W_i \text{ for } i \in S_t)$ \\
    \quad (e.g., Federated Averaging: \\$W_{\text{global}} = \sum_{i \in S_t} \frac{|D_i|}{\sum_{j \in S_t} |D_j|} W_i$)
\ENDFOR
\STATE \textbf{return} final global model $W_{\text{global}}$
\end{algorithmic}
\end{algorithm}

In practical applications of V2X and ISAC, each edge device may only be able to observe partial regions of spectrum coverage, which calls for 
efficient distributed learning equipped with 
multi-user semantic communications. However, radiomaps usually contain 
sensitive information, such as landscape distribution and user private information and patterns. To enable collaborative 
training of the cGAN backbone while avoiding privacy leakage, we integrate 
federated learning (FL) \cite{McMahan2017} with physics-enhanced semantic communications for multi-user model training.



The extended 
FL-based framework is shown in Fig. \ref{fig:3}, where 
each client trains a local model, 
to reconstruct the local radiomap based on collected 
sparse observations. 
During each synchronization cycle, each client shares its 
locally trained model 
with the global server.
The global server aggregates these updates via a weighted averaging, where each client’s dataset size is taken into consideration for computing the aggregated global model parameters, ensuring that larger datasets have a proportionally greater influence on the global update. Different from traditional FedAvg \cite{McMahan2017}, our framework relies on the cGAN architecture, where Eq.~\eqref{CGAN} is used 
in model training. The training process of our FedCGAN for multi-user semantic communications are described in Algorithm 1.

\section{Experimental Results}
We evaluate
different radiomap transmission approaches by comparing  their performance and efficiency in reconstruction accuracy, bandwidth usage and downstream tasks.

\textbf{Dataset}: We use the RadioMapSeer dataset \cite{Yapar2024data} for performance evaluation, which is simulated by \textit{WinProp} with size $256\times 256$. In our experiments, we split the dataset of 400/100/100 samples for training/validation/testing, where each sample represents a radiomap data point containing BS locations, building segmentation, and signal power. To calculate the bandwidth usage of landscape images, online resource such as \textit{OpenStreetMap} or \textit{Google Maps} are utilized to scratch the image samples with size of $256\times 256$.
\begin{table}[t] 
    \centering 
    \caption{Test MSE ($\times 10^{-4}$) Comparison For sample-free RME} 
    \vspace{-2mm}
    \begin{tabular}{c|c} 
    \hline 
        Method  & MSE(E-4)\\ 
    \hline 
    \hline
        AE & 210.6612\\
        RU & 19.4825\\ 
        cGAN & 94.6961\\
        RME-GAN & 99.9894 \\
       Geo2SigMap &  8.6920\\
        TiRE-GAN & 7.9489 \\
        Our method & \underline{6.6587} \\
    \hline 
    \end{tabular} 
    \label{tab:NMSEandMSE} 
    \vspace{-5mm} 
\end{table}
\subsection{Performance of cGAN-Backbone in Sample-Free RME}
As illustrated in Section \ref{backbone}, the cGAN backbone at the receiver end has a similar function as sample-free radiomap estimation from environmental information, where $0\%$ ratio of sparse observations are considered. To validate the efficiency of our physics-enhanced cGAN, we compare our backbone with SOTA RME approaches, include AutoEncoder (AE) \cite{Yapar2024data}, RadioUnet (RU) \cite{Levie2021}, traditional cGAN \cite{Mirza2014}, RME-GAN \cite{RMEGAN2023}, TiRE-GAN \cite{Tire} and Geo2SigMap \cite{GEO2SIG}.
All the models are trained for 120 epochs with a learning rate of 1e-4. We use the Adam optimizer with a batch size of 32. The MSE performance of different methods is compared 
in Table \ref{tab:NMSEandMSE}. From the results, our physics-enhanced cGAN always outperform 
the other benchmarks, 
thanks to the physics-enhanced utilization of radio depth map to embed the radio propagation information. Since RME-GAN requires some samples to embed radio propagation models and balance the sampling regions, it degrades to traditional cGAN in the sample-free scenarios, which leads to a significant performance decrease compared to non-zero sample ratio scenarios. The results verify the advantages of our backbone models, 
leading to 
better radiomap reconstruction accuracy in semantic communications.


\vspace{-2mm}
\subsection{Trade-off of Bandwidth Usage vs. Reconstruction Accuracy}
\vspace{-1mm}

\begin{figure}[t]
    \centering
    \includegraphics[width=0.95\linewidth]{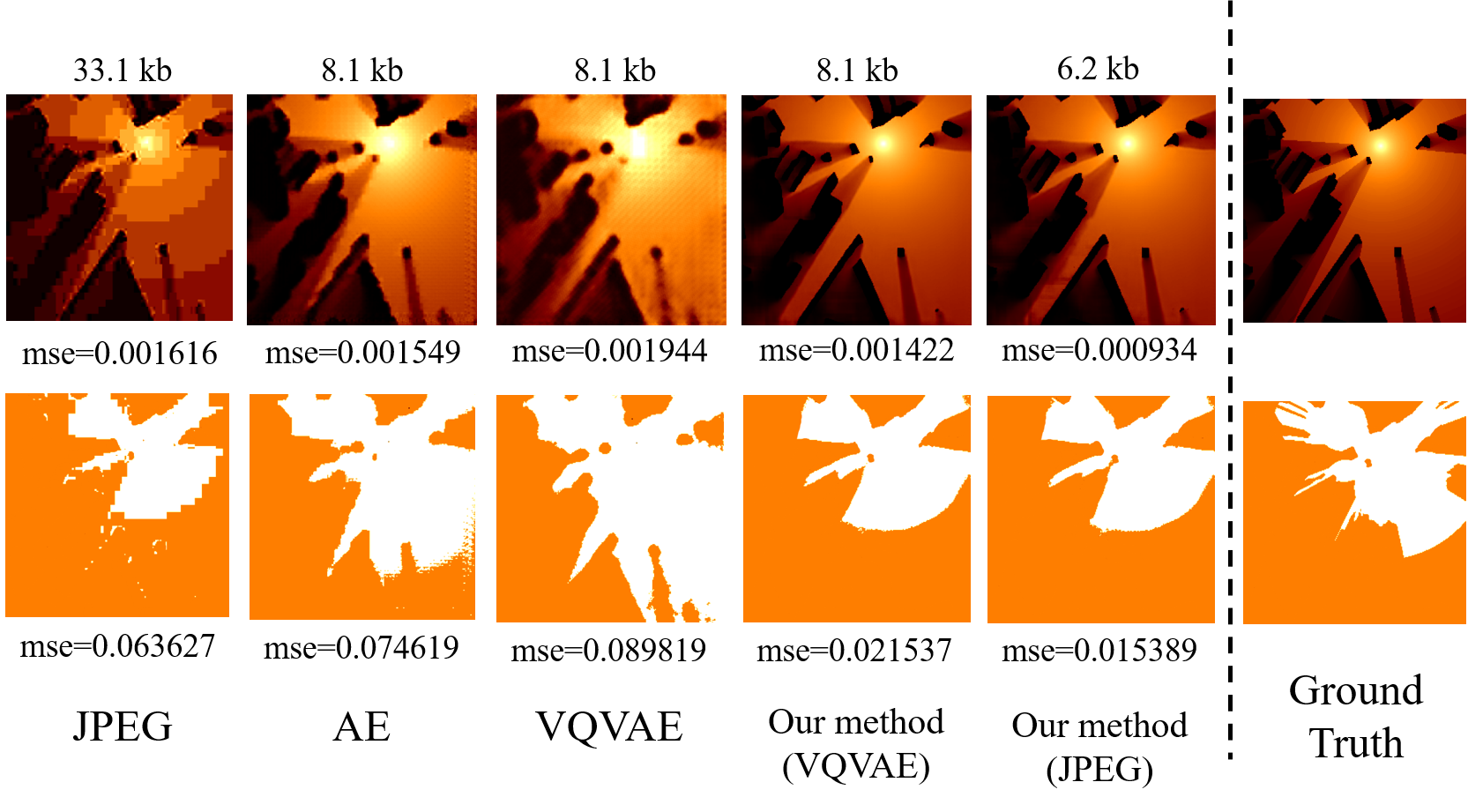}
    \vspace{-4mm}
    \caption{Results of RME and outage prediction (outage areas marked as orange) under similar bandwidth usage.}
    \label{fig:visualization}
    \vspace{-4mm}
\end{figure}

\begin{figure}[t]
    \centering
    \includegraphics[width=0.62\linewidth]{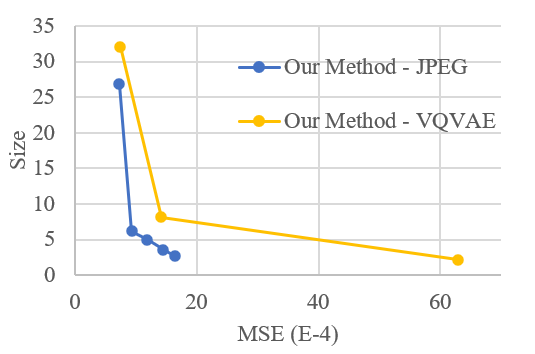}
    \vspace{-3mm}
    \caption{Ablation study of different semantic compression methods.}
    \vspace{-3mm}
    \label{fig:4-2}
\end{figure}

Next, we present the bandwidth usage and reconstruction accuracy of radiomap transmission compared to other existing 
learning-based communication frameworks, including those based on AE \cite{Bourtsoulatze2019}, variational autoencoder (VAE) \cite{Saidutta2021} and VQVAE \cite{VQVAE}. We also compare the results with JPEG compression. We first compare the bandwidth usage for radiomap transmission under similar reconstruction performance. As shown in Table \ref{tab:my_label}, our method could significantly reduce the communication overhead compared to data-oriented communications and other learning-based frameworks. Particularly, since the JPEG-encoded binary building segmentation map is less sensitive to noise and loss, our method with JPEG-based semantic compression has better 
performance than VQVAE. 
It is worth noting that the VQVAE-based semantic communications could achieve desired performance as well, but at the cost of sufficiently large training samples, which is not always the real case in wireless applications. 
In addition, we also present the results of outage prediction and RME under comparable 
bandwidth usage as shown in Fig. \ref{fig:visualization}. 
The outage prediction is 
measured by a binary classifier 
with a pre-trained neural networks. The result show that 
our 
physics-enhanced semantic communications outperforms other approaches, which is consistent with Table \ref{tab:my_label}. We also present the results of ablation study of different semantic compression schemes 
under our framework in Fig. \ref{fig:4-2}, where JPEG works better with limited bandwidth while VQVAE is preferred when 
low-error tolerance. 

\begin{table}[t]
    \centering
    \caption{Performance of RM Reconstruction and Bandwidth Usage}
    \begin{tabular}{c|cc}
    \hline
       \textbf{method}  &  \textbf{MSE(E-4)} & \textbf{Bandwidth(kb)}\\
       \hline
       \hline
     Raw Data & -- & 59.1\\
        JPEG & 490.15 & 29.8\\
       AE & 15.50 & 8.1\\
       VQVAE & 19.44 & 8.1\\
       VAE & 77.53& 32.1\\
         Our method (VQVAE) & \underline{14.22} & 8.1\\
       Our method (JPEG)&14.53& \underline{3.6}\\
       \hline
       
    \end{tabular}
    
    \label{tab:my_label}
    \vspace{-3mm}
\end{table}

\subsection{FL for Multi-user Semantic Communications}
We now evaluate the performance of the FL-based multi-user semantic communications compared to the centralized training and individual training counterparts.  In this paper, 
we consider 4 clients, and each client contains 8000/2000/2000 samples for training/validation/testing, respectively. For the centralized training, we assume the cloud server could observe all the raw data, which provides 
the best performance but the worst data privacy. For the individual training, each client train its own model from the local data without collaboration with others. For our FedCGAN, the client-server pairs follow Algorithm 1 for collaborative model training without sharing the raw data. The results are shown in Fig. \ref{table:results_global}, where the individual training performs the worst due to the lack of global data distributions. Our proposed FedCGAN achieves the desired 
performance in MSE/NMSE, which is 
comparable to the centralized training, while preserving the privacy. 

    
\begin{table}[t]
    \centering
    \caption{MSE and NMSE values for Multi-user Training}
    \renewcommand{\arraystretch}{1.5}
    \begin{tabular}{c|cccc}
        \hline
        & Individual & FedCGAN (proposed) & Centralized\\
        \hline
        \hline
        \textbf{MSE (E-4)}  & 13.305 & 6.32 & 6.66 \\
        \textbf{NMSE} & 1.145 & 1.16 & 1.03 \\
        \hline
    \end{tabular}
    \label{table:results_global}
    \vspace{-3mm}
\end{table}

\section{Conclusion}
In this work, we propose a novel physics-enhanced semantic communications of radio coverage information transmission for radiomap-assisted applications. Instead of bit-wise data-oriented message transmission, we 
embed the radio propagation behavior by the most critic semantics of radiomaps, including the LDPL model parameters and surrounding environments. By utilizing 
semantic compression via VQVAE and JPEG, our method 
significantly reduces the communication overhead for radio coverage information transmission. With a pre-trained physics-enhanced cGAN, 
dense radiomaps can be accurately recovered at the receiver for the radiomap-assisted spectrum management. 
Experimental results in 
radiomap recovery and outage map reconstruction verify  
the efficiency and effectiveness of our proposed framework in reducing the communication overhead and increasing the RME accuracy. 


\end{document}